\pdfoutput=1
\documentclass[11pt,letterpaper]{article}
\usepackage[margin=1in]{geometry}
\usepackage{amsmath,amssymb,amsthm,mathtools}
\usepackage{enumitem}
\usepackage{xcolor}
\usepackage[font=small,skip=4pt]{caption}
\usepackage{tikz}
\usepackage{titlesec}
\titlespacing*{\section}{0pt}{2.2ex plus .5ex minus .2ex}{1ex}
\titlespacing*{\subsection}{0pt}{1.8ex plus .5ex minus .2ex}{.8ex}
\titlespacing*{\paragraph}{0pt}{1ex plus .3ex minus .2ex}{1em}
\usetikzlibrary{decorations.pathreplacing,patterns}
\usepackage[colorlinks=true,linkcolor=blue!50!black,citecolor=blue!50!black,urlcolor=blue!50!black]{hyperref}
\hypersetup{pdftitle={Tight B-Trees for Adjacency Labeling},pdfauthor={Noy Rotbart}}

\newtheorem{theorem}{Theorem}[section]
\newtheorem{lemma}[theorem]{Lemma}
\newtheorem{corollary}[theorem]{Corollary}
\newtheorem{proposition}[theorem]{Proposition}

\newcommand{\sP}{\boxtimes}
\DeclareMathOperator{\lca}{lca}
\DeclareMathOperator{\nid}{nid}
\DeclareMathOperator{\pth}{path}
\newcommand{\R}{\mathcal R}

\title{Tight B-Trees for Adjacency Labeling}
\author{Noy Rotbart\\ {\small Churney, Copenhagen, Denmark}\\ {\small\texttt{noy@churney.io}}}
\date{}

\begin{document}
\maketitle

\begin{abstract}
We show that planar graphs admit adjacency labels of size $\log n+O(\log\log n)$: every vertex gets a label of this many bits, and the adjacency of two vertices is decided from their labels alone, in constant time. This improves the $\log n+O(\sqrt{\log n\log\log n})$ of Dujmovi\'c et al.\ (J.~ACM~2021) and the $\log n+O(\sqrt{\log n})$ of Gawrychowski and Janczewski (SOSA~2022). Equivalently, \mbox{$n$-vertex} planar graphs have an induced-universal graph with $n(\log n)^{O(1)}$ vertices, down from $n2^{O(\sqrt{\log n})}$. The same holds for graphs of bounded genus, apex-minor-free graphs and $k$-planar graphs. For every proper minor-closed class we get $\log n+O(\sqrt{\log n\log\log n})$, down from the $\log n+O((\log n)^{3/4})$ of Dujmovi\'c et al.\ (2026). Both results come from \emph{tight B-trees}: nodes are split into parts of nearly equal size, so every node but the topmost stays within a factor $1\pm O(1/\log n)$ of its target size, and a root-to-leaf path loses only $O(\log\log n)$ bits.
\end{abstract}

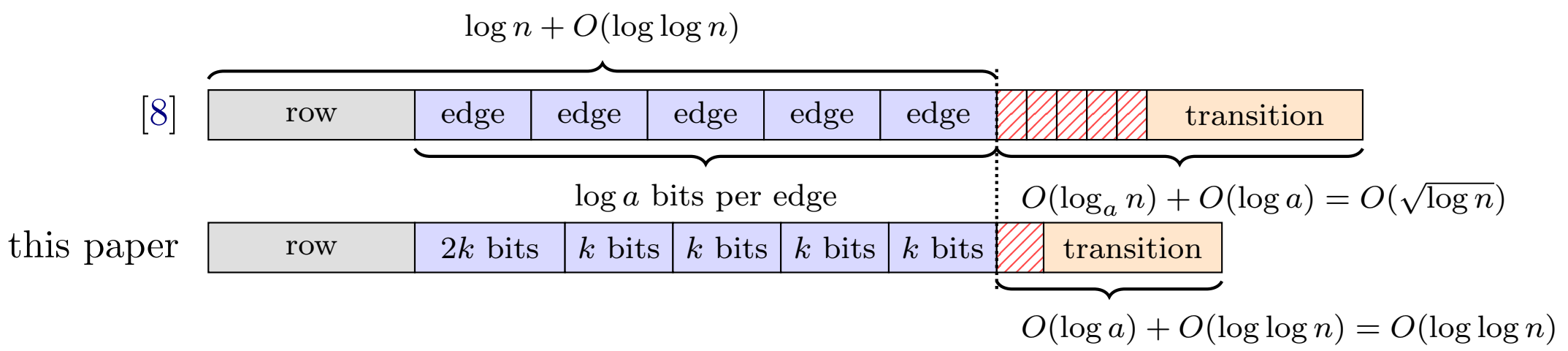
\begin{figure}[t]
\centering
\begin{tikzpicture}[x=0.85cm,y=1cm]
\def\h{0.45}
\begin{scope}[yshift=1.2cm]
\node[anchor=east] at (-0.15,\h/2) {\small\cite{GJ22}};
\draw[fill=gray!25] (0,0) rectangle (2.2,\h); \node at (1.1,\h/2) {\scriptsize row};
\foreach \i in {0,...,4}{
  \draw[fill=blue!15] (2.2+1.24*\i,0) rectangle (3.44+1.24*\i,\h); \node at (2.82+1.24*\i,\h/2) {\scriptsize edge};
  \draw[pattern=north east lines,pattern color=red!70] (8.4+0.32*\i,0) rectangle (8.72+0.32*\i,\h);
}
\draw[fill=orange!20] (10.0,0) rectangle (12.3,\h); \node at (11.15,\h/2) {\scriptsize transition};
\draw[decorate,decoration={brace,amplitude=4pt,mirror},thick] (2.2,-0.08)--(8.4,-0.08) node[midway,below=5pt] {\scriptsize $\log a$ bits per edge};
\draw[decorate,decoration={brace,amplitude=4pt,mirror},thick] (8.4,-0.08)--(12.3,-0.08);
\node[anchor=north west] at (8.5,-0.24) {\scriptsize $O(\log_an)+O(\log a)=O(\sqrt{\log n})$};
\end{scope}
\begin{scope}[yshift=0cm]
\node[anchor=east] at (-0.15,\h/2) {\small this paper};
\draw[fill=gray!25] (0,0) rectangle (2.2,\h); \node at (1.1,\h/2) {\scriptsize row};
\draw[fill=blue!15] (2.2,0) rectangle (3.8,\h); \node at (3.0,\h/2) {\scriptsize $2k$ bits};
\foreach \i in {0,...,3}{
  \draw[fill=blue!15] (3.8+1.15*\i,0) rectangle (4.95+1.15*\i,\h); \node at (4.375+1.15*\i,\h/2) {\scriptsize $k$ bits};
}
\draw[pattern=north east lines,pattern color=red!70] (8.4,0) rectangle (8.9,\h);
\draw[fill=orange!20] (8.9,0) rectangle (10.8,\h); \node at (9.85,\h/2) {\scriptsize transition};
\draw[decorate,decoration={brace,amplitude=4pt,mirror},thick] (8.4,-0.08)--(10.8,-0.08);
\node[anchor=north west] at (8.5,-0.24) {\scriptsize $O(\log a)+O(\log\log n)=O(\log\log n)$};
\end{scope}
\draw[densely dotted,thick] (8.4,-0.15)--(8.4,1.85);
\draw[decorate,decoration={brace,amplitude=4pt},thick] (0,1.75)--(8.4,1.75) node[midway,above=5pt] {\scriptsize $\log n+O(\log\log n)$};
\end{tikzpicture}
\caption{Composition of a label: the row code (gray) and the path without slack (blue) take $\log n+O(\log\log n)$ bits; then the slack in the balance of the trees (red), and the transition label and other fields (orange). Here $a=2^{\sqrt{\log n}}$ in~\cite{GJ22}, and $a=\Theta(\log^2n)$ in this paper.}\label{fig:label}
\end{figure}

\section{Introduction}

An \emph{adjacency labeling scheme} for a family $\mathcal F$ of graphs assigns a binary label to every vertex of a graph $G\in\mathcal F$, such that a decoder can determine whether $uv\in E(G)$ from the labels of $u$ and $v$ alone. The \emph{size} of the scheme is the maximum label length over the $n$-vertex graphs of $\mathcal F$. Kannan, Naor and Rudich~\cite{KNR92} noticed that a labeling scheme of size $f(n)$ is equivalent to an \emph{induced-universal graph} with $2^{f(n)}$ vertices, that is, a graph that contains every $n$-vertex graph of $\mathcal F$ as an induced subgraph. Since the label size is in the exponent, every bit counts: size $\log n+O(1)$ corresponds to an induced-universal graph with $O(n)$ vertices.

\subsection{Previous Work}

Kannan et al.~\cite{KNR92} gave labels of size $4\log n$ for planar graphs. In 2007, Gavoille and Labourel~\cite{GL07} improved this to $(2+o(1))\log n$, for every proper minor-closed class. In 2020 Bonamy, Gavoille and Pilipczuk~\cite{BGP20} reached $(4/3+o(1))\log n$ for planar graphs, and shortly after, Dujmovi\'c et al.~\cite{DEGJMM21} gave the first scheme of size $(1+o(1))\log n$. Their scheme builds on the product structure theorem~\cite{DJMMUW20}: every planar graph is a subgraph of $H\sP P$, where $H$ has bounded treewidth and $P$ is a path. Its vertices therefore lie in rows. Within a row, a vertex is found by a root-to-leaf path in a balanced search tree over the row, as in~\cite{GL07}, and to reach the next row it stores a short \emph{transition label}, from which the decoder computes its path in the tree of that row. Hence the trees of consecutive rows must be similar, and the resulting slack in their balance is paid on every path. In~\cite{DEGJMM21} the trees are binary, and balancing their height against the length of the transition labels gives size $\log n+O(\sqrt{\log n\log\log n})$. In 2022, Gawrychowski and Janczewski~\cite{GJ22} replaced the binary trees by B-trees with branching parameter $a$, rebuilt one level per row. A node of level $h$ in their trees has $\Theta(a^h)$ leaves, and so a path pays $O(1)$ bits of slack on each of its at most $\log_an$ edges, while the other parts of a label cost $O(\log a+\log\log n)$ bits. Choosing $a=2^{\sqrt{\log n}}$ gives size $\log n+O(\sqrt{\log n})$, with a constant-time decoder. Very recently, Dujmovi\'c et al.~\cite{DGJMMW26} reached $\log n+O((\log n)^{3/4})$ for every proper minor-closed class, by combining labeling schemes along tree-decompositions whose torsos have product structure.

\subsection{Our Results}

Our results are the following.

\begin{theorem}\label{thm:main}
For $1\le t\le\log n$, the family of $n$-vertex subgraphs of $H\sP P$, where $H$ has treewidth at most $t$ and $P$ is a path, admits a labeling scheme of size $\log n+O(t\log\log n)$.
\end{theorem}

\begin{corollary}\label{cor:planar}
Planar graphs admit a labeling scheme of size $\log n+O(\log\log n)$. Equivalently, some graph with $n(\log n)^{O(1)}$ vertices contains all $n$-vertex planar graphs as induced subgraphs. The same holds for graphs of bounded genus, apex-minor-free graphs and $k$-planar graphs for fixed $k$.
\end{corollary}

\begin{theorem}\label{thm:minor}
Every proper minor-closed class admits labels of size $\log n+O(\sqrt{\log n\log\log n})$. Equivalently, it has induced-universal graphs with $n2^{O(\sqrt{\log n\log\log n})}$ vertices.
\end{theorem}

Hence the smallest induced-universal graph for planar graphs is now known up to a polylogarithmic factor. To our knowledge, no lower bound of the form $\log n+\omega(1)$ is known, and whether planar graphs admit labels of size $\log n+O(1)$, as trees do~\cite{ADK17}, remains open.

\paragraph{Our methods.} As in~\cite{GJ22}, we build B-trees $T_1,\dots,T_m$ over sets $V_1,\dots,V_m$ in which consecutive sets differ in few elements of every interval (Lemma~\ref{lem:cascade}), and obtain $T_{j+1}$ from $T_j$ by rebuilding one level. We rebuild a level by cutting the grandchildren of every node into groups of nearly equal size. A new node of level $h$ is then one of $c\ge a/3$ nearly equal parts, and so has $(1\pm O(1/a))a^h$ leaves. Its level is rebuilt again within $\log n$ steps, and until then its number of leaves changes only by a factor $1\pm O(1/\log n)$ (Lemma~\ref{lem:inv} and Figure~\ref{fig:core}). With $a$ slightly below a power of two $2^k$, a path of $k$-bit child indices is then only $O(k)$ bits longer than the logarithm of the number of leaves (Lemma~\ref{lem:code}), and it changes in three indices from one tree to the next (Lemma~\ref{lem:trans}). As $a$ and the cascading parameter are polylogarithmic, every overhead is $O(\log\log n)$ bits; see Figure~\ref{fig:label}.

For Theorem~\ref{thm:minor} we use the framework of~\cite{DGJMMW26}, which needs a \emph{weighted mixed} labeling scheme for the class $\R_t$ of subgraphs of $H\sP P$ with $H$ of treewidth at most $t$. There, a vertex $x$ of weight $\omega(x)$ gets $\log\omega(G)-\log\omega(x)+g_1$ bits, a clique $K$ gets $\log\omega(G)-\log\min_K\omega+g_3$ bits, and the framework loses $O(\sqrt{g_3\log n}+\sqrt{\log n/g_3}\,\log\log n)$ bits. The trees of~\cite{GJ22} give $g_3=\Theta(\sqrt{\log n})$. In ours, a vertex of weight $\omega(x)$ needs only $\log|V_j|-\log\omega(x)+O(\log\log n)$ bits of a path, and so $g_1,g_3=O(\log\log n)$ (Lemma~\ref{lem:weighted}). Since this loss is $\Omega(\sqrt{\log n\log\log n})$ for every $g_3$, Theorem~\ref{thm:minor} is the limit of the framework as analyzed in~\cite{DGJMMW26}.

\section{Preliminaries}\label{sec:prelim}

All logarithms are binary, and we write $x=(1\pm\varepsilon)y$ if $|x-y|\le\varepsilon y$. For a graph $H$ and a path $P$ with vertices $1,\dots,m$ in order, two distinct vertices $(v,i)$ and $(v',j)$ of $H\sP P$ are adjacent if $v=v'$ or $vv'\in E(H)$, and $|i-j|\le 1$. We call $V(H)\times\{i\}$ \emph{row} $i$. Every planar graph is a subgraph of $H\sP P$ for a graph $H$ of treewidth at most $8$ and a path $P$~\cite{DJMMUW20}.

Every graph of treewidth at most $t$ is a spanning subgraph of a $t$-tree $H'$. We fix a perfect elimination ordering of $H'$, in which the neighbors of every vertex $v$ that precede it form a clique, and let $C(v)$ consist of $v$ and these neighbors. Note that $C(v)$ is a clique, $|C(v)|\le t+1$, and every clique $Q$ of $H'$ satisfies $Q\subseteq C(v)$ for its last vertex $v$. We use the following two tools of Gawrychowski and Janczewski~\cite{GJ22}, the first of which builds on~\cite{GL07}.

\begin{proposition}\label{prop:int}
For every $t$-tree $H'$ on at most $n$ vertices there is a map $v\mapsto I(v)=[\ell(v),r(v)]$ to integer intervals such that adjacent vertices get intersecting intervals, at most $\theta=O(t\log n)$ of the intervals pairwise intersect, and the right endpoints $r(v)$ are distinct.
\end{proposition}

\begin{lemma}\label{lem:row}
For positive integers $w_1,\dots,w_m$ with sum $N$ there are prefix-free strings $\rho_1,\dots,\rho_m$ with $|\rho_i|\le\log(N/w_i)+3$, and strings $\sigma_i,\pi_i$ of $O(\log\log N)$ bits, such that a fixed function $f$ satisfies $f(\rho_i,\rho_j,\sigma_i,\pi_j)=1$ if and only if $j=i+1$.
\end{lemma}

The next two lemmas are variants of lemmas in~\cite{GJ22}. We prove them, since we apply them to other trees and with $q$ growing with $n$; the proof of Lemma~\ref{lem:cascade} is in Appendix~\ref{app:cascade}. Let $T$ be an ordered rooted tree some of whose leaves carry the elements of a set $V$ of integers, in increasing order from left to right. For $u\in V(H')$ with $I(u)\cap V\ne\emptyset$, we denote by $\lca(u,T)$ the lowest common ancestor of the leaves that carry $I(u)\cap V$.

\begin{lemma}\label{lem:lca}
A node of $T$ with $\Delta$ children is $\lca(u,T)$ for at most $(\Delta+1)\theta$ vertices $u$. If $Q$ is a clique of $H'$ and $b=\min\{r(u):u\in Q\}\in V$, then the leaf $b$ lies below $\lca(u,T)$ for all $u\in Q$.
\end{lemma}
\begin{proof}
Intervals with a common point pairwise intersect, and so at most $\theta$ of them contain a given integer. If $\lca(u,T)$ is a leaf, then $I(u)$ contains its integer. Otherwise $I(u)$ contains two consecutive elements of $V$ that are carried below different children of $\lca(u,T)$, and below a node with $\Delta$ children there are at most $\Delta-1$ such pairs. For the second claim, the intervals of the clique $Q$ pairwise intersect, and so $\ell(u)\le b\le r(u)$ for every $u\in Q$.
\end{proof}

\begin{lemma}\label{lem:cascade}
For finite sets of integers $Y_1,\dots,Y_m$ with $\sum_i|Y_i|=M$ and an integer $q\ge 1$, there are sets $Y_i\subseteq V_i\subseteq\bigcup_jY_j$ with $\sum_i|V_i|\le2q(M+m)$ such that for every $j<m$ and every interval $J$ of integers, $|(V_j\setminus V_{j+1})\cap J|\le\lceil|V_j\cap J|/q\rceil$ and $|(V_{j+1}\setminus V_j)\cap J|\le\lceil|V_{j+1}\cap J|/q\rceil$.
\end{lemma}

\section{Tight B-Trees}\label{sec:trees}

In this section $V_1,\dots,V_m$ are sets of integers with the property of Lemma~\ref{lem:cascade}, for the $q$ below, and $1\le|V_j|\le n$. Nothing in this section refers to graphs. We set
\begin{equation}\label{eq:param}
\Lambda=\lceil\log n\rceil,\qquad \varepsilon=\frac{1}{8\Lambda},\qquad q=64\Lambda^2,\qquad k=\lceil\log q\rceil+1,\qquad a=(1-3\varepsilon)2^k,
\end{equation}
so that $64\Lambda^2\le a<2^k<4q$. We assume that $n\ge16$; smaller $n$ only affect the constants.

\begin{lemma}\label{lem:drift}
Let $J$ be an interval and $x_i=|V_i\cap J|$. Then $x_{j+s}=(1\pm\varepsilon/4)x_j\pm 2\Lambda$ for $0\le s\le\Lambda$.
\end{lemma}
\begin{proof}
By Lemma~\ref{lem:cascade}, $x_{j+1}\ge x_j-\lceil x_j/q\rceil\ge(1-1/q)x_j-1$, and symmetrically $x_j\ge(1-1/q)x_{j+1}-1$. Iterating gives $(1-1/q)^sx_j-s\le x_{j+s}\le(1-1/q)^{-s}(x_j+s)$, and since $s\le\Lambda\le q/2$, we have $(1-1/q)^{-s}\le1+2s/q\le1+\varepsilon/4$ by~\eqref{eq:param}.
\end{proof}

\paragraph{Trees.} A \emph{row tree} is an ordered rooted tree in which the root has level $\Lambda$, and every node of level $h\ge2$ has children, all of level $h-1$. Every node $x$ carries an interval $J_x$ of integers, such that the root carries $\mathbb Z$ and the intervals of the children of a node partition its interval, in order. Given a set $V$ of integers, the \emph{leaves} of a node $x$ of level $1$ are the elements of $V\cap J_x$, in order; we view them as level $0$. It follows that every node $x$ has $w(x)=|V\cap J_x|$ leaves. We call the least level $h\ge1$ with a single node the \emph{height} $L$, and its node the \emph{apex}. Since every node of level $h\ge2$ has a child, the levels above $L$ also have a single node.

\paragraph{Regrouping.} To \emph{regroup} level $p\in\{1,\dots,\Lambda-1\}$, we do the following for every node $y$ of level $p+1$. Let $w=w(y)$, and let $c=\lfloor w/a^p+1/2\rfloor$ if $w\ge a^{p+1}/2$, and $c=1$ otherwise. We cut the sequence of grandchildren of $y$ into $c$ groups, placing the $j$-th cut after the first grandchild for which the grandchildren up to it have a total number of leaves closest to $jw/c$. The children of $y$ are then replaced by $c$ new nodes, one per group, whose children are the grandchildren in their group and whose intervals partition $J_y$, in order. (For $p=1$ the grandchildren are leaves, and the intervals are chosen to contain them.)

\paragraph{The procedure.} The tree $T_1$ for $V_1$ is built top-down: every node of level $h+1\ge2$ with $w\ge a^{h+1}/2$ leaves gets $\lfloor w/a^h+1/2\rfloor$ children whose numbers of leaves differ by at most one, and every other node gets a single child. For $j\ge1$, we obtain $T_{j+1}$ from $T_j$ by replacing $V_j$ with $V_{j+1}$, and then regrouping level $p=1+((j-1)\bmod(\Lambda-1))$. Thus every level is regrouped once in every $\Lambda-1$ consecutive steps. Note that a node keeps its interval for as long as it exists, and therefore has $|V_i\cap J_x|$ leaves in every tree $T_i$ that contains it; Lemma~\ref{lem:drift} applies to it.

\begin{figure}[t]
\centering
\begin{tikzpicture}[x=1cm,y=1cm]
\begin{scope}
\draw[fill=blue!12,draw=black!45] (0,0) rectangle (0.239,0.32);
\draw[fill=blue!12,draw=black!45] (0.239,0) rectangle (0.488,0.32);
\draw[fill=blue!12,draw=black!45] (0.488,0) rectangle (0.767,0.32);
\draw[fill=blue!12,draw=black!45] (0.767,0) rectangle (1.007,0.32);
\draw[fill=blue!12,draw=black!45] (1.007,0) rectangle (1.251,0.32);
\draw[fill=blue!12,draw=black!45] (1.251,0) rectangle (1.502,0.32);
\draw[fill=blue!12,draw=black!45] (1.502,0) rectangle (1.719,0.32);
\draw[fill=blue!12,draw=black!45] (1.719,0) rectangle (1.963,0.32);
\draw[fill=blue!12,draw=black!45] (1.963,0) rectangle (2.217,0.32);
\draw[fill=blue!12,draw=black!45] (2.217,0) rectangle (2.485,0.32);
\draw[fill=blue!12,draw=black!45] (2.485,0) rectangle (2.695,0.32);
\draw[fill=blue!12,draw=black!45] (2.695,0) rectangle (2.921,0.32);
\draw[fill=blue!12,draw=black!45] (2.921,0) rectangle (3.129,0.32);
\draw[fill=blue!12,draw=black!45] (3.129,0) rectangle (3.399,0.32);
\draw[fill=blue!12,draw=black!45] (3.399,0) rectangle (3.658,0.32);
\draw[fill=blue!12,draw=black!45] (3.658,0) rectangle (3.862,0.32);
\draw[fill=blue!12,draw=black!45] (3.862,0) rectangle (4.147,0.32);
\draw[fill=blue!12,draw=black!45] (4.147,0) rectangle (4.43,0.32);
\draw[fill=blue!12,draw=black!45] (4.43,0) rectangle (4.686,0.32);
\draw[fill=blue!12,draw=black!45] (4.686,0) rectangle (4.939,0.32);
\draw[fill=blue!12,draw=black!45] (4.939,0) rectangle (5.154,0.32);
\draw[fill=blue!12,draw=black!45] (5.154,0) rectangle (5.356,0.32);
\draw[fill=blue!12,draw=black!45] (5.356,0) rectangle (5.602,0.32);
\draw[fill=blue!12,draw=black!45] (5.602,0) rectangle (5.808,0.32);
\draw[fill=blue!12,draw=black!45] (5.808,0) rectangle (6.025,0.32);
\draw[fill=blue!12,draw=black!45] (6.025,0) rectangle (6.246,0.32);
\draw[fill=blue!12,draw=black!45] (6.246,0) rectangle (6.449,0.32);
\draw[fill=blue!12,draw=black!45] (6.449,0) rectangle (6.689,0.32);
\draw[fill=blue!12,draw=black!45] (6.689,0) rectangle (6.927,0.32);
\draw[fill=blue!12,draw=black!45] (6.927,0) rectangle (7.199,0.32);
\draw[densely dashed,red!70!black] (1.8,-0.05)--(1.8,0.62);
\draw[densely dashed,red!70!black] (3.599,-0.05)--(3.599,0.62);
\draw[densely dashed,red!70!black] (5.399,-0.05)--(5.399,0.62);
\draw[very thick] (1.719,-0.12)--(1.719,0.44);
\draw[very thick] (3.658,-0.12)--(3.658,0.44);
\draw[very thick] (5.356,-0.12)--(5.356,0.44);
\draw[decorate,decoration={brace,amplitude=4pt},thick] (0.03,0.72)--(1.689,0.72) node[midway,above=4pt] {\scriptsize $(1\pm\frac3a)a^p$};
\draw[decorate,decoration={brace,amplitude=4pt},thick] (1.749,0.72)--(3.628,0.72) node[midway,above=4pt] {\scriptsize $(1\pm\frac3a)a^p$};
\draw[decorate,decoration={brace,amplitude=4pt},thick] (3.688,0.72)--(5.326,0.72) node[midway,above=4pt] {\scriptsize $(1\pm\frac3a)a^p$};
\draw[decorate,decoration={brace,amplitude=4pt},thick] (5.386,0.72)--(7.169,0.72) node[midway,above=4pt] {\scriptsize $(1\pm\frac3a)a^p$};
\node[anchor=north,red!70!black] at (1.8,-0.06) {\scriptsize $w/c$};
\node[anchor=north,red!70!black] at (3.599,-0.06) {\scriptsize $2w/c$};
\node[anchor=north,red!70!black] at (5.399,-0.06) {\scriptsize $3w/c$};
\node[anchor=north] at (3.599,-0.4) {\scriptsize grandchildren of $y$, at most $1.2a^{p-1}$ leaves each};
\node[anchor=east] at (-0.1,0.16) {\small (a)};
\end{scope}
\begin{scope}[xshift=8.2cm]
\fill[blue!12] (0,0.15) rectangle (5.6,0.95);
\draw[black!60] (0,0.15)--(5.6,0.15) (0,0.95)--(5.6,0.95);
\draw[densely dashed,black!45] (0,-0.2)--(5.6,-0.2) (0,1.3)--(5.6,1.3);
\draw[black!40] (0,0.55)--(5.6,0.55);
\node[anchor=west] at (5.65,0.95) {\scriptsize $(1+\varepsilon)a^h$};
\node[anchor=west] at (5.65,0.15) {\scriptsize $(1-\varepsilon)a^h$};
\node[anchor=west] at (5.65,0.55) {\scriptsize $a^h$};
\node[anchor=west,black!60] at (5.65,1.3) {\scriptsize $\Theta(a^h)$ in~\cite{GJ22}};
\draw[thick,blue!60!black] (0.0,0.6)--(0.174,0.617)--(0.348,0.651)--(0.523,0.692)--(0.697,0.754)--(0.871,0.788)--(1.045,0.82)--(1.219,0.754)--(1.394,0.749)--(1.568,0.811)--(1.742,0.82)--(1.916,0.82)--(2.09,0.766)--(2.265,0.762)--(2.439,0.726)--(2.613,0.732)--(2.787,0.743)--(2.961,0.674)--(3.135,0.635)--(3.31,0.604)--(3.484,0.662)--(3.658,0.699)--(3.832,0.652)--(4.006,0.693)--(4.181,0.643)--(4.355,0.659)--(4.529,0.607)--(4.703,0.537)--(4.877,0.589)--(5.052,0.548)--(5.226,0.509)--(5.4,0.576);
\fill (0,0.6) circle (1.6pt);
\draw[densely dashed] (5.4,-0.3)--(5.4,1.4);
\node[anchor=north] at (0,-0.32) {\scriptsize created};
\node[anchor=north] at (5.4,-0.32) {\scriptsize replaced, $<\Lambda$ rows later};
\node[anchor=east] at (-0.1,0.55) {\small (b)};
\end{scope}
\end{tikzpicture}
\caption{The core idea. (a) Regrouping cuts closest to $jw/c$, so every new child has $(1\pm3/a)a^p$ leaves. (b)~Until replaced, a node of level $h$ keeps $(1\pm\varepsilon)a^h$ leaves (Lemma~\ref{lem:inv}); in~\cite{GJ22}, only $\Theta(a^h)$.}\label{fig:core}
\end{figure}
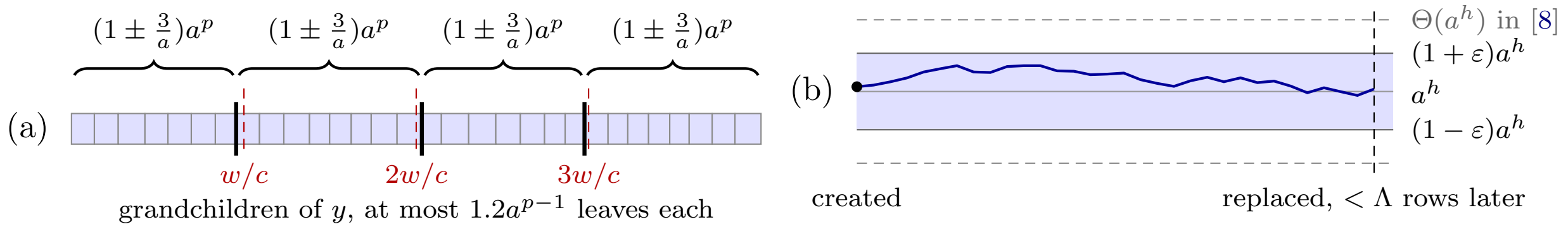

The next lemma is the heart of the paper; Figure~\ref{fig:core} illustrates it.

\begin{lemma}\label{lem:inv}
Let $L$ be the height of $T_j$. Then $|V_j|\le a^{L+1}$, and every node of level $h<L$ in $T_j$ has $(1\pm\varepsilon)a^h$ leaves, with respect to $V_j$ and, if $j<m$, also with respect to $V_{j+1}$.
\end{lemma}
\begin{proof}
We call a node \emph{split} if it was created together with a sibling, in $T_1$ or by a regroup. We show by induction on $j$ that in $T_j$ every node of level $1\le h<L$ is split, that every split node of level $h$ had $(1\pm3/a)a^h$ leaves when it was created, and that $|V_j|\le a^{L+1}$. This suffices, since a split node of $T_j$ was created fewer than $\Lambda-1$ steps earlier, and by Lemma~\ref{lem:drift} has $(1\pm\varepsilon/4)(1\pm3/a)a^h\pm2\Lambda=(1\pm\varepsilon)a^h$ leaves with respect to $V_j$ and $V_{j+1}$, using $3/a\le\varepsilon/4$ and $2\Lambda\le\varepsilon a/4$.

We start with an observation. Suppose that a regroup gives a node $y$ of level $p+1$ with $w$ leaves $c\ge2$ children, and that every grandchild of $y$ has at most $1.2a^{p-1}$ leaves. Then $w\ge a^{p+1}/2$, and so $c\ge a/3$ and $|w/c-a^p|\le a^p/(2c)\le1.5a^{p-1}$. Each cut is within $0.6a^{p-1}$ of its target, and it follows that every child of $y$ has $w/c\pm1.2a^{p-1}=(1\pm3/a)a^p$ leaves. The same holds in $T_1$, where the children of $y$ have $w/c\pm1$ leaves.

Consider first $T_1$. Its root has a single child, as $|V_1|\le n<a^\Lambda/2$, so $L<\Lambda$, and $|V_1|<a^{L+1}/2$ as the node above the apex has a single child. If $L\ge2$, the nodes of level $L-1$ are at least two children of the apex, and by the observation every node of level $h+1<L$ has at least $a^{h+1}/2$ leaves, and hence at least two children. Thus every node of level $1\le h<L$ is split.

Consider next the regroup of level $p$ that gives $T_{j+1}$, and let $L$ be the height of $T_j$ and $W=|V_{j+1}|$. By induction and Lemma~\ref{lem:drift}, $W\le(1+\varepsilon/4)a^{L+1}+2\Lambda\le(1+\varepsilon/2)a^{L+1}$. Moreover, if $p\le L$, the grandchildren of the nodes of level $p+1$ have level $p-1<L$, and therefore at most $(1+\varepsilon)a^{p-1}$ leaves each. We distinguish four cases.
\begin{itemize}[itemsep=0pt,topsep=1pt,partopsep=0pt,parsep=0pt,leftmargin=1.2em]
\item $p\le L-2$. Every node $y$ of level $p+1$ is split, so $w(y)\ge(1-\varepsilon)a^{p+1}$, and $y$ gets $c\ge2$ new children. The height stays $L$.
\item $p=L-1$. If $W\ge a^L/2$, the apex gets $c\ge2$ new children and the height stays $L$. Otherwise its single new child becomes the apex, and the height drops to $L-1$, since $W\ge2(1-\varepsilon)a^{L-1}$ exceeds the number of leaves of any node of level $L-2$.
\item $p=L$. If $W\ge a^{L+1}/2$, the node above the apex gets $c\ge2$ new children and becomes the apex, and the height rises to $L+1$. Otherwise the height stays $L$.
\item $p>L$. A node of level $p+1$ has $W<a^{p+1}/2$ leaves and a single grandchild, so its child is replaced by an identical node.
\end{itemize}
Thus nodes below the new height are split, and new ones have $(1\pm3/a)a^p$ leaves by the observation.

It remains to bound $|V_j|$. While the height is $L$, every regroup of level $L$ finds fewer than $a^{L+1}/2$ leaves. When the height becomes $L$, there are fewer than $a^{L+1}/2$ leaves if it comes from $L+1$ or in $T_1$, and by induction at most $(1+\varepsilon/2)a^L<a^{L+1}/2$ leaves if it comes from $L-1$. The last such event happened fewer than $\Lambda-1$ steps before step $j$, and so Lemma~\ref{lem:drift} with $J=\mathbb Z$ gives $|V_j|\le(1+\varepsilon/4)a^{L+1}/2+2\Lambda\le a^{L+1}$.
\end{proof}

Lemma~\ref{lem:inv} also bounds the numbers of children. A node of level $h<L$ has at most $(1+\varepsilon)a/(1-\varepsilon)<2^k$ children. If $L\ge2$, the apex has at least two children, so $|V_j|\ge2(1-\varepsilon)a^{L-1}$, and hence $L\le1+\log n/\log a<\Lambda$. The apex has at most $a^{L+1}/((1-\varepsilon)a^{L-1})<2^{2k}$ children.

\paragraph{Path codes.} For a leaf $b$ of $T_j$ and $1\le h\le L$, let $d_h$ be the index, counted from $0$, of the child taken at the ancestor of $b$ of level $h$. Then $d_h<2^k$ for $h<L$, and $d_L<2^{2k}$. We let $\pth(b,T_j)$ be the string of $(L+1)k$ bits that writes $d_L$ with $2k$ bits, followed by $d_{L-1},\dots,d_1$ with $k$ bits each. The ancestor of $b$ of level $h<L$ is then identified by $h$ and the first $(L+1-h)k$ bits of $\pth(b,T_j)$, and the apex by its level.

\begin{lemma}\label{lem:code}
Let $L$ be the height of $T_j$. Then $(L-1)k\le\log|V_j|$, and so $|\pth(b,T_j)|\le\log|V_j|+2k$ for every leaf $b$ of $T_j$.
\end{lemma}
\begin{proof}
The first claim is clear for $L=1$. For $L\ge2$, note that $1-3\varepsilon\ge2^{-6\varepsilon}$, and so $\log a\ge k-6\varepsilon$. Therefore $\log|V_j|\ge 1+\log(1-\varepsilon)+(L-1)(k-6\varepsilon)\ge(L-1)k$, using $\log(1-\varepsilon)\ge-1/4$ and $6\varepsilon L\le3/4$. The second claim follows, as $|\pth(b,T_j)|=(L+1)k$.
\end{proof}

\begin{lemma}\label{lem:trans}
For $b\in V_j\cap V_{j+1}$ there is a string $\delta(b,j)$ of $O(k+\log\Lambda)$ bits such that $\pth(b,T_{j+1})$ and the height of $T_{j+1}$ are determined by $\pth(b,T_j)$, the height of $T_j$ and $\delta(b,j)$.
\end{lemma}
\begin{proof}
Replacing $V_j$ with $V_{j+1}$ changes only $d_1$, and regrouping level $p$ changes only $d_p$ and $d_{p+1}$. Therefore $\delta(b,j)$ stores $p$, the new height, and the new $d_1$, $d_p$ and $d_{p+1}$.
\end{proof}

\section{The Labeling Scheme}\label{sec:scheme}

We now prove Theorem~\ref{thm:main}. Let $G\subseteq H\sP P$ have $n$ vertices, where $H$ has treewidth at most $t\le\log n$ and $P$ has vertices $1,\dots,m$. We may assume that $|V(H)|\le n$ and that no row is empty, since contracting an edge of $P$ at an empty row keeps $G\subseteq H\sP P$. We fix a $t$-tree $H'\supseteq H$ with sets $C(v)$ and intervals $I(v)$ as in Proposition~\ref{prop:int}. Let $S_i=\{v:(v,i)\in V(G)\}$, $X_i=\{r(u):u\in C(v),\,v\in S_i\}$ and $X_0=\emptyset$, and apply Lemma~\ref{lem:cascade} with $q$ as in~\eqref{eq:param} to the sets $Y_i=X_{i-1}\cup X_i$. The resulting sets $V_i$ consist of right endpoints, so $1\le|V_i|\le n$, and $\sum_i|V_i|=O(qtn)$ as $\sum_i|Y_i|\le2(t+1)n$ and $m\le n$. On them we build the trees of Section~\ref{sec:trees}.

For $u$ with $r(u)\in V_i$, let $\nid(u,T_i)$ be the rank of $u$ among the vertices with the same node $\lca(\cdot,T_i)$. As every node has fewer than $2^{2k}$ children, it takes at most $(2^{2k}+1)\theta$ values by Lemma~\ref{lem:lca}. The \emph{identity} $(\lca(u,T_i),\nid(u,T_i))$ of $u$ in $T_i$ determines $u$.

\paragraph{Labels.} For $x=(v,i)\in V(G)$ let $b=\min\{r(u):u\in C(v)\}$ and $C(v)=\{u_0,\dots,u_z\}$ with $u_0=v$. Then $b\in X_i$ lies in $V_i$, and also in $V_{i+1}$ if $i<m$. The label of $x$ consists of (i) the strings $\rho_i,\sigma_i,\pi_i$ of Lemma~\ref{lem:row} for the weights $|V_i|+1$, the height of $T_i$, and $z$; (ii) $\pth(b,T_i)$, and for every $u_e$ the level of $\lca(u_e,T_i)$ and $\nid(u_e,T_i)$; (iii) $\delta(b,i)$, and for every $u_e$ the level of $\lca(u_e,T_{i+1})$ and $\nid(u_e,T_{i+1})$, or zeros if $i=m$; and (iv) for every $u_e$ and $\tau\in\{-1,0,1\}$, a bit telling whether $x$ is adjacent in $G$ to $(u_e,i+\tau)$. By Lemma~\ref{lem:lca} with $Q=C(v)$ and by Lemma~\ref{lem:trans}, the label gives the identities of the $u_e$ in $T_i$ and, if $i<m$, in $T_{i+1}$.

\paragraph{Decoding.} Given the labels of $x=(v,i)$ and $x'=(v',j)$, the decoder decides by Lemma~\ref{lem:row} whether $i=j$ or $|i-j|=1$, and outputs $0$ otherwise; we may assume $j\in\{i,i+1\}$. It computes the identities of both lists in $T_j$. If $v'=u_e$ for some $u_e$ in the list of $x$, with $e\ge 1$ if $i=j$, it outputs the bit $(e,j-i)$ of $x$; if $v$ is in the list of $x'$, it outputs the corresponding bit of $x'$; and otherwise it outputs $0$. If $xx'\in E(G)$, then either $v=v'$ and $i\ne j$, so $v'=u_0$, or $vv'\in E(H')$, so $v'\in C(v)$ or $v\in C(v')$. As identities determine vertices, the decoder outputs $1$ exactly when $xx'\in E(G)$.

\paragraph{Size.} Every label starts with $\lceil\log n\rceil$ in a self-delimiting code, and every field is preceded by its length, in $O(t\log\log n)$ bits in total. For $N=\sum_i(|V_i|+1)=O(qtn)$, Lemmas~\ref{lem:row} and~\ref{lem:code} give $|\rho_i|+|\pth(b,T_i)|\le\log N+2k+3$. Since $k,\log q,\log t,\log\Lambda=O(\log\log n)$ and $\theta=O(t\log n)$, the label has size $\log n+O(t\log\log n)$. This proves Theorem~\ref{thm:main}. Corollary~\ref{cor:planar} follows by the product structure theorems of~\cite{DJMMUW20,DMW23} and by~\cite{KNR92}.

\paragraph{Running time.} The labels can be computed in polynomial time~\cite{DJMMUW20}. As in~\cite{GJ22}, on a word RAM with $\Theta(\log n)$-bit words, each step of the decoder takes constant time: reading a field, rewriting a path by Lemma~\ref{lem:trans}, comparing identities, and evaluating $f$. Hence the decoder runs in $O(t)$ time.

\bibliographystyle{plain}

\appendix
\section{Proof of Lemma~\ref{lem:cascade}}\label{app:cascade}

\begin{proof}[Proof of Lemma~\ref{lem:cascade}]
For a finite set $Z$ of integers, let $Z^\circ$ be $Z$ without its elements whose rank in $Z$, counted from $1$, is a multiple of $q$. Let $F_1=Y_1$ and $F_{i+1}=Y_{i+1}\cup F_i^\circ$, let $B_m=Y_m$ and $B_i=Y_i\cup B_{i+1}^\circ$, and let $V_i=F_i\cup B_i$. Consider $x\in V_j\setminus V_{j+1}$. Clearly $x\in F_j$, since otherwise $x\in B_j\setminus Y_j\subseteq B_{j+1}$. Moreover $x\notin F_{j+1}\supseteq F_j^\circ$, and so the rank of $x$ in $F_j$ is a multiple of $q$. Since $F_j\cap J$ consists of at most $|V_j\cap J|$ elements of consecutive ranks in $F_j$, the first inequality follows, and the second is symmetric. Finally, $|F_{i+1}|\le|Y_{i+1}|+|F_i|-\lfloor|F_i|/q\rfloor\le|Y_{i+1}|+(1-1/q)|F_i|+1$, and summing over $i$ gives $\sum_i|F_i|\le q(M+m)$. The same holds for the sets $B_i$.
\end{proof}

\section{Proof of Theorem~\ref{thm:minor}}\label{app:minor}

We use the following definition of~\cite[Section~3]{DGJMMW26}. A graph class $\mathcal G$ admits a \emph{weighted $(g_1,g_2,g_3)$ mixed labeling scheme} if there are functions $A$ and $I$ such that for every $n$-vertex graph $G^+\in\mathcal G$, every spanning subgraph $G$ of $G^+$ and every $\omega:V(G)\to\mathbb R_{>0}$, there are strings $\mu(x)$ for $x\in V(G)$, $\mu(K)$ for the cliques $K$ of $G^+$, and $\kappa(K,x)$ for $x\in K$, such that (i)~$\mu(K)\ne\mu(K')$ for distinct cliques $K$ and $K'$; (ii)~$A(\mu(x),\mu(y))=1$ if and only if $xy\in E(G)$; (iii)~$I(\mu(K),\kappa(K,x),\mu(y))=1$ if and only if $x=y$; (iv)~$|\mu(x)|\le\log\omega(G)-\log\omega(x)+g_1(n)$; (v)~$|\mu(K)|\le\log\omega(G)-\log\min_{x\in K}\omega(x)+g_3(n)$; and (vi)~$|\kappa(K,x)|\le g_2(n)$, where $\omega(G)=\sum_{x\in V(G)}\omega(x)$. With $\omega\equiv1$, (ii) and (iv) give a labeling scheme of size $\log n+g_1(n)$. Let $\R_t$ be the class of graphs that are subgraphs of $H\sP P$ for a graph $H$ of treewidth at most $t$ and a path $P$.

\begin{lemma}\label{lem:weighted}
For every fixed $t$, the class $\R_t$ admits a weighted $(g_1,g_2,g_3)$ mixed labeling scheme with $g_1(n),g_3(n)=O(\log\log n)$ and $g_2(n)=O(1)$.
\end{lemma}

\paragraph{Proof of Lemma~\ref{lem:weighted}.} Let $G^+$, $G$ and $\omega$ be as above, where $G^+\subseteq H\sP P$ and $H$ has treewidth at most $t$. As in Section~\ref{sec:scheme}, we may assume that $|V(H)|\le n$, that $P$ has vertices $1,\dots,m$ and no row of $G^+$ is empty, and we fix $H'$, the sets $C(v)$ and the intervals $I(v)$. As $G^+\subseteq H'\sP P$, every clique of $G^+$ lies in two consecutive rows and projects to a clique of $H'$. Let $\omega'(x)=\lceil n\omega(x)/\omega(G)\rceil$. Then $\omega'$ takes positive integer values, $\omega'(G)\le2n$, and $\log n-\log\omega'(x)\le\log\omega(G)-\log\omega(x)$. It therefore suffices to give vertex labels of size $\log n-\log\omega'(x)+O(\log\log n)$ and clique labels of size $\log n-\log\min_{x\in K}\omega'(x)+O(\log\log n)$.

\emph{Copies.} For $u\in V(H')$ and $1\le i\le m$, let $\nu(u,i)$ be the sum of $\omega'(v,j)$ over the vertices $(v,j)$ of $G$ with $j\in\{i-1,i\}$ and $u\in C(v)$. Let $Y_i$ consist of the pairs $(r(u),c)$ with $1\le c\le\nu(u,i)$, which we call \emph{copies} of $r(u)$, ordered lexicographically and identified with integers in the same order. Since $\nu(u,j)$ and $\nu(u,j+1)$ both count $\omega'(v,j)$, we have the following property~(P): \emph{for every vertex $(v,j)$ of $G$ and $u\in C(v)$, the first $\omega'(v,j)$ copies of $r(u)$ belong to $Y_j$, and also to $Y_{j+1}$ if $j<m$.} Moreover $\sum_i|Y_i|=\sum_{u,i}\nu(u,i)\le2(t+1)\omega'(G)\le4(t+1)n$. We apply Lemma~\ref{lem:cascade} to $Y_1,\dots,Y_m$ and build the trees $T_1,\dots,T_m$ of Section~\ref{sec:trees} on the resulting sets $V_i$, with $n$ replaced by $n'=4(t+1)n\ge|V_i|$ in~\eqref{eq:param} and in the first field of the labels. As $t$ is fixed, $k$, $\log q$ and $\log\Lambda$ are still $O(\log\log n)$, and $N=\sum_i(|V_i|+1)=O(qn)$. Lemma~\ref{lem:lca} and its proof hold verbatim if $I(u)\cap V$ denotes the set of copies of points in $I(u)$, with points in place of integers, and with any copy of $b$ in place of the leaf $b$. Identities are defined as in Section~\ref{sec:scheme}.

\emph{Truncated paths.} For a leaf $\beta$ of $T_j$ and a level $h$, let $\pth_h(\beta,T_j)$ be the first $(L+1-h)k$ bits of $\pth(\beta,T_j)$, where $L$ is the height of $T_j$, or the empty string if $h\ge L$. Together with $h$ and $L$, it identifies the ancestors of $\beta$ of level at least $h$. For $\beta\in V_j\cap V_{j+1}$, the proof of Lemma~\ref{lem:trans} shows that $\pth_h(\beta,T_{j+1})$ and the height of $T_{j+1}$ are determined by $\pth_h(\beta,T_j)$, $h$, $L$ and $\delta(\beta,j)$: the digits $d_g$ with $g>h$ are copied from $\pth_h(\beta,T_j)$, except those that $\delta(\beta,j)$ replaces.

\emph{Levels.} If $Y_j$ contains $s$ copies of $r(u)$, they are leaves below $\lca(u,T_j)$. By Lemma~\ref{lem:inv}, a node of level $h$ of $T_j$ has at most $(1+\varepsilon)a^h$ leaves if $h<L$, and at most $|V_j|\le a^{L+1}$ if $h=L$. So if $h$ is the level of $\lca(u,T_j)$, then $s\le2a^{h+1}$, and hence $hk\ge h\log a\ge\log s-k-1$. Now let $\beta$ be a leaf of $T_j$, let $1\le s\le|V_j|$, and let $h$ be a level with $hk\ge\log s-k-1$. Then Lemma~\ref{lem:code} gives the bound~(T): $|\pth_h(\beta,T_j)|\le\max\{0,(L+1-h)k\}\le\log|V_j|-\log s+3k+1$.

\emph{Vertex labels.} For $x=(v,i)\in V(G)$, let $b=\min\{r(u):u\in C(v)\}$ and $\beta=(b,1)$. By~(P), $\beta\in V_i$, and $\beta\in V_{i+1}$ if $i<m$. Let $h$ be the least level of the nodes $\lca(u,T_i)$ and, if $i<m$, $\lca(u,T_{i+1})$, over $u\in C(v)$. The label $\mu(x)$ is the label of Section~\ref{sec:scheme}, with $\pth(b,T_i)$ and $\delta(b,i)$ replaced by $h$, $\pth_h(\beta,T_i)$ and $\delta(\beta,i)$. By Lemma~\ref{lem:lca}, the nodes $\lca(u,T_i)$ for $u\in C(v)$ are ancestors of $\beta$ of level at least $h$, and so are the nodes $\lca(u,T_{i+1})$ in $T_{i+1}$. By the paragraph on truncated paths, $\mu(x)$ therefore gives the identities of all $u\in C(v)$ in $T_i$ and $T_{i+1}$, and the decoder of Section~\ref{sec:scheme}, which uses only these identities and the bits~(iv), serves as $A$. By~(P), $Y_i$ and $Y_{i+1}$ contain $\omega'(x)$ copies of $r(u)$ for every $u\in C(v)$, and so $hk\ge\log\omega'(x)-k-1$. Now~(T) and Lemma~\ref{lem:row} give $|\rho_i|+|\pth_h(\beta,T_i)|\le\log N-\log\omega'(x)+3k+4$. The other fields take $O(\log\log n)$ bits as in Section~\ref{sec:scheme}, and $\log N=\log n+O(\log\log n)$. Hence $|\mu(x)|\le\log n-\log\omega'(x)+O(\log\log n)$.

\emph{Clique labels.} Let $K$ be a clique of $G^+$, let $i$ be the largest row that meets $K$, and let $Q=\{u:(u,j)\in K\}$. Then $K$ lies in rows $i-1$ and $i$, $Q$ is a clique of $H'$, and $|K|\le2|Q|\le2t+2$. For $(u,j)\in K$, (P) applied to $(u,j)$ and $u\in C(u)$ shows that $Y_i$ contains $\omega'(u,j)$ copies of $r(u)$. Let $b=\min\{r(u):u\in Q\}$, so that $\beta=(b,1)\in V_i$, and let $h$ be the least level of the nodes $\lca(u,T_i)$, $u\in Q$. The label $\mu(K)$ consists of $\rho_i,\sigma_i,\pi_i$, the height of $T_i$, $h$, $\pth_h(\beta,T_i)$, and, for every $(u,j)\in K$ in a fixed order, the bit $i-j$, the level of $\lca(u,T_i)$ and $\nid(u,T_i)$, each field preceded by its length. We let $\kappa(K,(u,j))$ be the position of $(u,j)$ in this order, written with $\lceil\log(2t+2)\rceil$ bits. By Lemma~\ref{lem:lca}, $\mu(K)$ gives the identities in $T_i$ of all $u\in Q$, and hence determines $K$, which proves~(i). As $h$ is the level of $\lca(u,T_i)$ for some $(u,j)\in K$, we have $hk\ge\log\min_{y\in K}\omega'(y)-k-1$, and~(T) gives $|\mu(K)|\le\log n-\log\min_{y\in K}\omega'(y)+O(\log\log n)$, as for vertex labels.

\emph{Identity tester.} Given $\mu(K)$, $\kappa(K,(u,j))$ and $\mu(x)$ for $x=(v,i')$, the tester $I$ finds the entry of $(u,j)$ in $\mu(K)$, and hence $j\in\{i-1,i\}$. If $j=i$, it tests whether $\rho_{i'}=\rho_i$, and if $j=i-1$, whether $f(\rho_{i'},\rho_i,\sigma_{i'},\pi_i)=1$. If the test fails, then $i'\ne j$, and $I$ outputs $0$. Otherwise $i'=j$, and $\mu(x)$ gives the identity of $v$ in $T_i$, which is $T_{i'}$ or $T_{i'+1}$. The tester outputs $1$ if and only if it equals the identity of $u$ in $T_i$, that is, if and only if $x=(u,j)$. This proves~(iii) and the lemma, with $g_2=\lceil\log(2t+2)\rceil$. \qed

\paragraph{Proof of Theorem~\ref{thm:minor}.} We repeat the proof of Lemma~18 of~\cite{DGJMMW26}, with their Lemma~19 replaced by Lemma~\ref{lem:weighted}. Let $\mathcal C$ be a proper minor-closed class. By the graph minor product structure theorem~\cite{DJMMUW20} (see~\cite[Theorem~3]{DGJMMW26}), there are integers $t,s\ge0$ such that every graph in $\mathcal C$ has a tree-decomposition whose torsos are in $\R_t^{+s}$, the class of graphs $G$ with $G-X\in\R_t$ for some set $X$ of at most $s$ vertices. By Lemma~\ref{lem:weighted} and Lemmas~13 and~14 of~\cite{DGJMMW26}, the class $\mathcal Q$ of disjoint unions of graphs in $\R_t^{+s}$ admits a weighted $(g_1,g_2,g_3)$ mixed labeling scheme with $g_1,g_3=O(\log\log n)$ and $g_2=O(1)$, where we may take $g_3(n)=\Theta(\log\log n)$ and all $g_i$ non-decreasing. The class $\mathcal Q$ is hereditary and closed under disjoint unions, and its cliques have at most $2t+s+2$ vertices. Every graph in $\mathcal C$ therefore has a tree-decomposition of adhesion-width at most $2t+s+2$ whose torsos are in $\mathcal Q$. By Lemma~17 of~\cite{DGJMMW26}, $\mathcal C$ admits a weighted mixed labeling scheme with
\[
g_1''(n)=g_1(n)+O(g_2(n))+O\big(\sqrt{g_3(n)\log n}\big)+O\Big(\sqrt{\log n/g_3(n)}\,\log\log n\Big)=O\big(\sqrt{\log n\log\log n}\big),
\]
and hence a labeling scheme of size $\log n+O(\sqrt{\log n\log\log n})$. The universal graphs follow by~\cite{KNR92}. \qed

\end{document}